# FA-GAN: Fused Attentive Generative Adversarial Networks for MRI Image Super-Resolution

Mingfeng Jiang[1*], Minghao Zhi[1], Liying Wei[1], Xiaocheng Yang[1], Jucheng Zhang[2], Yongming Li[3], Pin Wang[3], Jiahao Huang[4,5,6], and Guang Yang[5, 6*]

[1] School of Information Science and Technology, Zhejiang Sci-Tech University, Hangzhou 310018, China

[2] Department of Clinical Engineering, the Second Affiliated Hospital, School of Medicine, Zhejiang University, Hangzhou, 310019, China

[3] College of Communication Engineering, Chongqing University, Chongqing, China

[4] School of Optics and Photonics, Beijing Institute of Technology, Beijing, China

[5] Cardiovascular Research Centre, Royal Brompton Hospital, London, SW3 6NP, UK

[6] National Heart and Lung Institute, Imperial College London, London, SW7 2AZ, UK

[*]Address correspondence to:

Prof. Mingfeng Jiang

School of Information Science and Technology, Zhejiang Sci-Tech University, Hangzhou 310018, P.R. China.

Phone: +86-571-86843312, Fax: +86-571-86843576

Email: m.jiang@zstu.edu.cn

Dr. Guang Yang

National Heart and Lung Institute, Imperial College London, London, SW7 2AZ, UK.

Phone: +44-2075948161, Fax: +44-2075948162

Email: g.yang@imperial.ac.uk

# Abstract

High-resolution magnetic resonance images can provide fine-grained anatomical information, but acquiring such data requires a long scanning time. In this paper, a framework called the Fused Attentive Generative Adversarial Networks(FA-GAN) is proposed to generate the super- resolution MR image from low-resolution magnetic resonance images, which can reduce the scanning time effectively but with high resolution MR images. In the framework of the FA-GAN, the local fusion feature block, consisting of different three-pass networks by using different convolution kernels, is proposed to extract image features at different scales. And the global feature fusion module, including the channel attention module, the self-attention module, and the fusion operation, is designed to enhance the important features of the MR image. Moreover, the spectral normalization process is introduced to make the discriminator network stable. 40 sets of 3D magnetic resonance images (each set of images contains 256 slices) are used to train the network, and 10 sets of images are used to test the proposed method. The experimental results show that the PSNR and SSIM values of the super-resolution magnetic resonance image generated by the proposed FA-GAN method are higher than the state-of-the-art reconstruction methods.

# 1. Introduction

Image super-resolution refers to the reconstruction of high-resolution images from low-resolution images [1]. High resolution means that the pixels in the image are denser and can display more flexible details [2][3]. These details are very useful in practical applications, such as satellite imaging, medical imaging, etc, which can better identify targets and find important features in high-resolution images [4-6].

High-resolution (HR) MRI images can provide fine anatomical information, which is helpful for clinical diagnosis and accurate decision-making[7][8]. However, it not only requires expensive equipment but also requires a long scanning time, which brings challenges to image data acquisition. Therefore, further applications are limited by slow data acquiring and imaging speed[9][10].

The super-resolution (SR) is a technique to generate a high-resolution (HR) image from a single or a group of low-resolution (LR) images, which can improve the visibility of image details or restore image details [11-13]. Without changing hardware or scanning components, SR methods can significantly improve the spatial resolution of MRI[14-15]. Generally, there are three methods to implement image SR in MRI: interpolation-based, construction-based, and machine learning-based[16][17].

The interpolation-based SR techniques assume that the area in the LR image can be extended to the corresponding area by using a polynomial or an interpolation function with a priori smoothness [18][19]. The advantages of the interpolation-based super-resolution reconstruction algorithm are simplicity and high real-time performance; the disadvantage is that it is too simple to make full use of the prior information of MR images. In particular, the super-resolution reconstruction algorithm based on a single MR image has obvious shortcomings, which in a blurred version of the corresponding HR reference image [20][21].

The reconstruction-based SR methods are introduced to solve an optimization problem incorporating two terms: the fidelity term, which penalizes the difference between a degraded SR image and an observed LR image, and the regularization term, which promotes sparsity and inherent characteristics of recovering the SR signal[22][23].

The performance of these techniques becomes suboptimal especially in the high-frequency region when the input data becomes too sparse or the model becomes even slightly inaccurate[24-26]. These shortcomings reduce the effect of reconstruction-based SR methods to large magnifications, which may work well for small magnifications less than 4.

Machine learning techniques, particularly deep learning (DL)-based SR approaches, have recently attracted considerable attention because of their state-of-the-art performance in SR for natural images. Most recent algorithms rely on data-driven deep learning models to reconstruct the required details for accurate super-resolution [27][28]. Deep learning-based methods aim to automatically learn the relationship between input and output directly from the training samples [29][30]. At the same time, deep learning has also played a vital role in CT/PET image reconstruction, such as PET Image Reconstruction from Sinogram Domain[31-33].

With the development of deep learning, the Generative Adversarial Network (GAN) proposed by Goodfellow et al, has recently been demonstrated that it has good performance in image transformation and super-resolution imaging. Sanchez et al proposed the standard super-resolution GAN (SRGAN) framework for generating brain super-resolution images[34]. Most GAN-based image generation models are constructed using convolutional layers. Convolutions process information in local neighborhoods, however, using only convolutional layers is inefficient in establishing remote dependencies in images[35][36].

It is difficult to learn the dependencies between images using a small convolution kernel. However, the size of the convolution kernel is too large, which will reduce the model's performance. Besides, increasing the size of the convolution kernel can also expand the receptive field, but it inevitably increases the complexity of the model [37][38]. Zhang et al propose the Self-Attention Generative Adversarial Network (SAGAN) with attention-driven, long-range dependency modeling for image generation tasks[39].

In the previous work on reconstruction problems, deep learning based methods have two major issues[40]. Firstly, they treat each channel-wise feature equally, but contributions to the reconstruction task vary from different feature maps. Secondly, the receptive field in a convolutional layer may cause to lose contextual information from original images, especially those high-frequency components that contain valuable detailed information such as edges and texture. Therefore, the Channel-Attention module is designed to filter the useless features and to enhance the informative ones. Therefore, model parameters in shallower layers are to be updated mostly that are relevant to a given task. To the best of our knowledge, this is the first work to employ channel-wise attention to the MRI reconstruction problem[41][42]. Combining the idea of MR reconstruction and image super-resolution, some researchers work on recovering HR images from low-resolution under-sampled K-space data directly[43-46].

In this paper, a fused attentive generative adversarial network (FA-GAN) is proposed for generating super-resolution MR images from low-resolution MR ones. The novelty of this work can be concluded as following: 1)The local fusion feature block, consisting of different three-pass networks by using different convolution kernels, was proposed to extract image features at different scales, so as to improve the reconstruction performances of SR images; 2) The global feature fusion module, including the channel attention module, the self-attention module, and the fusion operation, was designed to enhance the important features of the MRI image, so that the super-resolution image is more realistic and closer to the original image; 3)The spectral normalization (SN) is introduced to the discriminator network, which can not only smooth and accelerate the training process of the deep neural network but also improve the model generalization performance.

## 2. Methodology

The proposed neural network model is designed to learn the image firstly, and then inversely map the LR image to the reference HR image[47][48]. This model only takes LR images as input to generate SR images. The operation can be defined as

$$I^{LR} = f(I^{HR}), \qquad (1)$$

$$I^{HR} = g(I^{LR}) = f^{-1}(I^{LR}) + R, \tag{2}$$

where $I^{LR}$, $I^{HR} \in \mathbb{R}^{m \times n}$ are respectively LR and HR MRI images of size $m \times n$ and $f: I^{HR} \in \mathbb{R}^{m \times n} \to I^{LR} \in \mathbb{R}^{m \times n}$ denotes the down-sampling process that creates a LR counterpart from an HR image.

### 2.1 SR Network with GAN

The network output is passed through a series of upsampling stages, where each stage doubles the input image size. The output is passed through a convolution stage to get the resolved image. Depending upon the desired scaling, the number of upsampling stages can be changed. The adversarial min-max problem is defined by

$$\min_G \max_D E_{I^{HR} \sim P_{train}(I^{HR})}[\log D(I^{HR})] + E_{I^{LR} \sim P_G(I^{LR})}[\log(1 - D(G(I^{LR})))]. \tag{3}$$

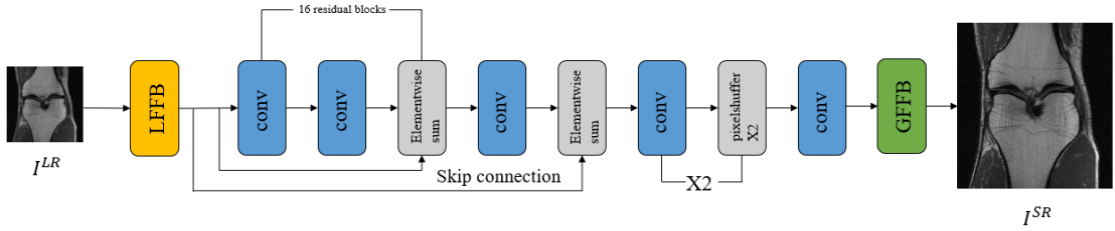

Figure 1: The framework of the proposed FA-GAN network. LFFB denotes local feature fusion block and GFFB denotes global feature fusion block.

The framework of the proposed FA-GAN network is shown in figure 1. The whole model takes the down-sampled low-resolution magnetic resonance image as input, extracts the features through the LFFB module, and generates the enlarged image through convolution and up-sampling. Finally, the GFFB module fuses the detailed features to generate a super-resolution magnetic resonance image. During the training process, HR references will be used to guide the optimization of model parameters. Moreover, the spectral normalization (SN) is introduced to the discriminator network to stabilize the training of GAN .

### 2.2 Local Fusion Feature Block (LFFB)

Different from those previous experiments[49], the local fusion feature block consists of different three-pass networks by using different convolution kernels, as shown in Figure 2. In this way, the information flows between those bypasses can be shared with each other, which allow our network to extract image features at different scales. The operation can be defined as

$$F_{t,1} = [C_{3\times3}(F_{t-1}), F_{t,1}], \tag{4}$$

$$F_{t,2} = [C_{5\times5}(F_{t-1}), F_{t,2}], \tag{5}$$

$$F_{t,3} = [C_{7\times7}(F_{t-1}), F_{t,3}], \tag{6}$$

$$F_d = F_{d,3} + F_{d-1}, \tag{7}$$

where $C_{s\times s}$ means the S scale feature extractor. Ours proposed S scale feature extractor consist of three convolution layers with $s\times s$ kernel size and one ReLU intermediate activation layer. The operation of F[ · ] means the concatenation and 1×1 convolution, which is mainly designed to quickly fuse features and reduce the computational burden.

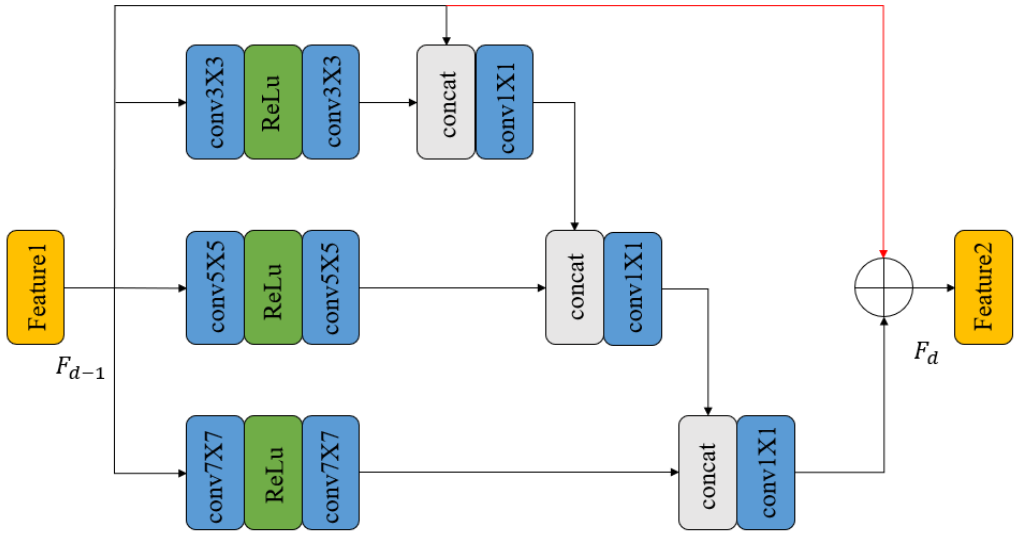

Figure 2: Local feature fusion block.

## 2.3 Global Feature Fusion Block （GFFB）

The global feature fusion module includes three parts, namely the channel attention module, the self-attention module, and the fusion operation. Through these modules, the important features of the MRI image can be enhanced, so that the super-resolution image is more realistic and closer to the original image.

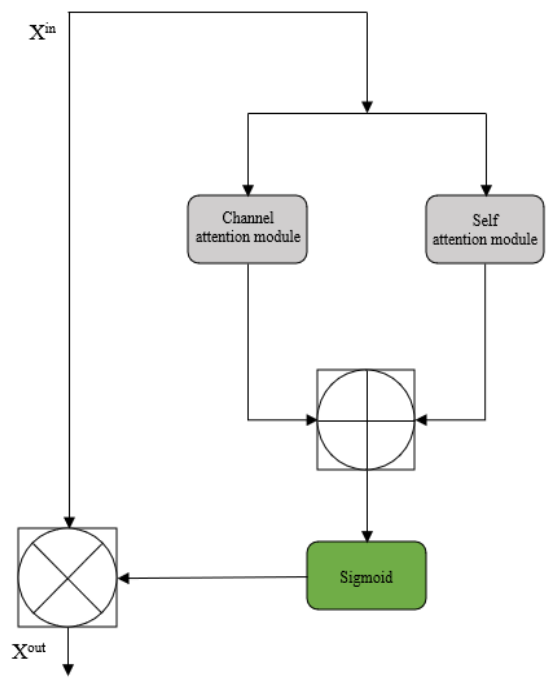

Figure 3: Global feature fusion block. GAP denotes the global average pooling operation.

**(1) Channel-Attention Module**

In this paper, a lightweight channel attention mechanism is introduced, which allows to selectively emphasize informative features and restrain less useful ones via a one-dimensional vector from global information. As illustrated in Figure 4, a global average pooling is used to extract the global information across spatial dimensions H*W firstly. Then, it is followed by a dimension reduction layer with a reduction ratio of $r$, a ReLu activation, a dimension increase layer, and a sigmoid activation to generate SR image. The two dimension computable layers are implemented by fully connected layers. The final output of the recalibration is acquired by rescaling the input features.

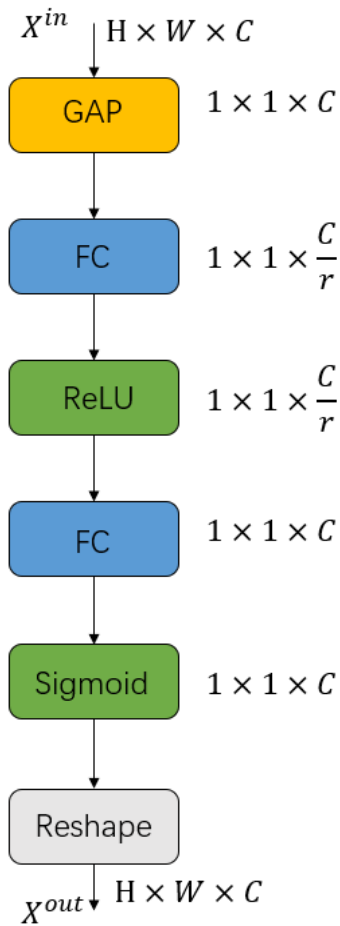

Figure 4: Channel-Attention module. GAP denotes the global average pooling operation and Reshape denotes enlarge operation.

**(2) Self-Attention Module**

The role of the self-attention module is to replace the traditional convolutional feature map with a self-attention feature map.

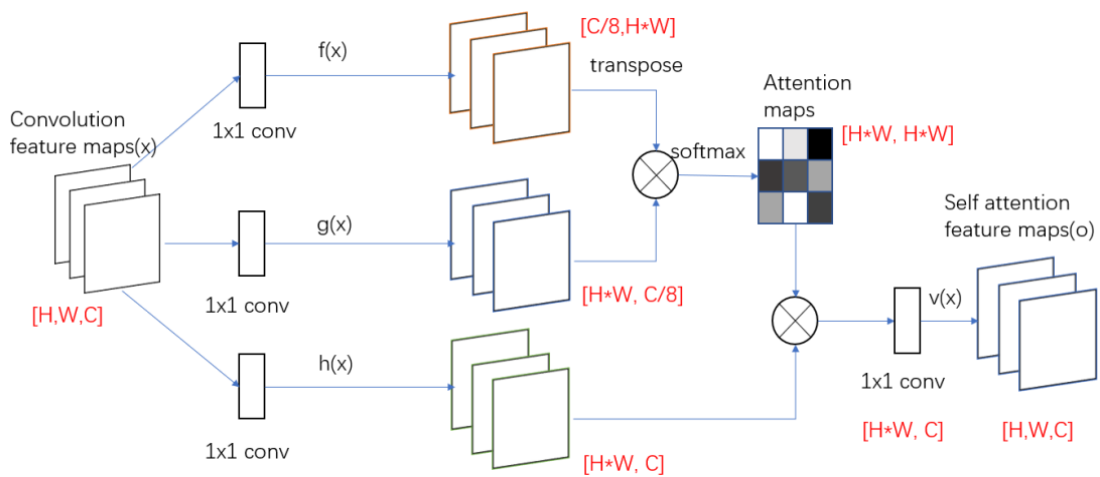

Figure 5: Self-Attention module.

After convolution operation, the convolutional feature maps pass three branches $f(x)$, $g(x)$, $h(x)$ of the 1x1 convolution structure, and the size of the feature map is unchanged. $g(x)$ changes the number of channels, and the output of $h(x)$ keeps the number of channels unchanged. H and W represent the length and width of the feature map, and C represents the number of channels. After transposing the output of $f(x)$, and multiplying the output matrix of $g(x)$, through normalizeing by softmax to get an [H*W, H*W] attention map. By multiplying the attention map with the output of $h(x)$ to get a [H*W, C] feature map, and using a 1x1 convolutions to reshape the output to [H, W, C] to get the feature map at this time. The structure of the self-attention module is shown in Figure 5.

$$s_{ij} = f(x_i)^T g(x_j) \tag{8}$$

$$\beta_{j,i} = \frac{\exp(s_{ij})}{\sum_{i=1}^{N} \exp(s_{ij})} \tag{9}$$

where $\beta_{j,i}$ indicates the extent to which the model attends to the $i^{th}$ location when synthesizing the $j^{th}$ region. Here, C is the number of channels and N is the number of feature locations of features from the previous hidden layer.

The output of the attention layer is $o$ and can be expressed as:

$$o_j = v(\sum_{i=1}^{N} \beta_{j,i} h(x_i)), \quad h(x_i) = W_h x_i, v(x_i) = W_v x_i. \tag{10}$$

In the above formulation, $w_g, w_f, w_h$, and $w_v$ are the learned weight matrices, which are implemented as 1×1 convolutions.

Besides, we further multiply the output of the attention layer by a scale parameter and add back the input feature map. Therefore, the final output is given by,

$$y_{SA} = \gamma o_i + x_i, \tag{11}$$

where γ is a learnable scalar and initialized to 0. Introducing learnable γ can make the network first rely on the information of the local neighborhood, and then gradually learn to assign more weight to non-local information.

**(3) Fusion Operation**

A. Direct Connection.

The direct connection function can be implemented by adding the two terms directly as following:

$$f(\alpha R_i^0, \beta Y_i) = \alpha R_i^0 + \beta Y_i \qquad (12)$$

where $i$ is the index of a feature. $R$ represents the output of Channel Attention, and $Y$ represents the output of Self-attention. Both $\alpha$ and $\beta$ are set to 0.5 as the preset value.

B. Weighted Connection.

Compared to the direct connection, the weighted connection introduces the competition between $R$ and $Y$. Besides, it can be easily extended to a softmax form, which is more robust and less sensitive to trivial features. Both α and β are set to 0.5 as the preset value. To avoid introducing extra parameters, we calculate weights using $R$ and $Y$. The weighted connection function is represented as

$$f(\alpha R_i^0, \beta Y_i) = \frac{\left|\alpha R_i^0\right|^2}{\alpha R_i^0 + \beta Y_i} + \frac{\left|\beta Y_i\right|^2}{\alpha R_i^0 + \beta Y_i} \qquad (13)$$

**(4) Loss Function**

The loss function is used to estimate the difference between the value generated or fitted by the model and the real value, that is, the difference between the reconstructed MRI and the original MRI. The smaller the loss function, the stronger the model is. In order to improve the quality of model reconstruction, we propose to use perceptual loss, pixel loss, and adversarial loss as the combined loss function of the generator. Perceptual loss mimics human visual differences, and pixel loss is the difference between pixels in the image domain.

$$l^{SR} = l_x^{SR} + 10^{-3} l_{Gen}^{SR} \qquad (14)$$

In the following we describe possible choices for the content loss $l_x^{SR}$ and the adversarial loss $l_{gen}^{SR}$.

This paper uses the Euclidean distance between VGG features, which is more

relevant to human perception, as the content loss, as shown below:

$$l_x^{SR} = \frac{1}{W_{i,j}H_{i,j}} \sum_{x=1}^{W_{i,j}} \sum_{y=1}^{H_{i,j}} (\phi_{i,j}(I^{HR})_{x,y} - \phi_{i,j}(G_{\theta_G}(I^{LR}))_{x,y})^2 \quad (15)$$

$\phi_{i,j}$ indicates that the extracted feature is the *j*-th convolutional layer before the *i*-th largest pooling layer. $W_{i,j}$, $H_{i,j}$ represents the dimension of the feature layer.

The adversarial loss function is the average discriminator probability value of the samples generated by the generator. The formula is as follows:

$$l_{Gen}^{SR} = \sum_{n=1}^{N} -\log D_{\theta_D}(G_{\theta_G}(I^{LR})) \quad (16)$$

$D_{\theta_D}(G_{\theta_G}(I^{LR}))$ represents the probability that the discriminator judges the image generated by the generator as the original magnetic resonance image.

## 3. Experimental Results

3.1 Datasets and metrics

All the experiment use TeslaV100-SXM2 GPU and four different MRI data sets to train and test the model. Randomly select 50 samples with 3D formats for training, of which 40 samples are used as the training set (3200 2D MRI) and 10 samples are used as validation set (960 2D MRI). All low-resolution images in the experiment are obtained by bicubic interpolation. At the same time, to ensure the fairness of the test, we conducted two independent tests to verify the performance of the proposed FA-GAN model. The first test experiment was to randomly select 10 samples for the test set (960 two-dimensional MRI images) to test and calculate the average quantitative index, and the second one was to select a two-dimensional MRI with obvious features from the testing sets. The optimizing procedure is implemented by using Adam optimization algorithm with 0.9. The FA-GAN networks were trained with a learning rate of 0.0001.The model training takes 10 hours at a time.

The experiment uses three evaluation criteria to evaluate the reconstructed image: peak signal-to-noise ratio (PSNR), structural similarity index measure (SSIM), and freshet Inception Distance score(FID). The definition of PSNR is

$$PSNR = 10 \times \log_{10}(\frac{255^2}{M \times N \sum_{i=1}^{M} \sum_{j=1}^{N}(y(i,j) - x(i,j))^2}), \quad (17)$$

where $x$ represents the original image, $y$ represents the super-resolution reconstructed image, and $i, j$ respectively represent the coordinate position of the pixel, and $M, N$ represent the size of the image. The SSIM can be defined by

$$SSIM = \frac{(2\mu_x\mu_y + C_1)(2\sigma_{xy} + C_2)}{(\mu_x^2 + \mu_y^2 + C_1)(\sigma_x^2 + \sigma_y^2 + C_2)}, \quad (18)$$

where $\mu_x$ and $\mu_y$ represent the mean of the image $x$ and $y$ respectively, $\sigma_x$ and $\sigma_y$ represent the variance of the image $x$ and $y$ respectively, the $\sigma_{xy}$ covariance of the image $x$ and $y$, $C_1$ and $C_2$ the constant value used to maintain stability. The expression of FID is

$$FID(x_r, g) = \| \mu_x - \mu_g \|_2^2 + Tr(\sum x_r + \sum g - 2(\sum x_r \sum g)^{\frac{1}{2}}) \quad (19)$$

In the formula, $Tr$ represents the sum of the elements on the diagonal of the matrix, $\mu$ is the mean, $\Sigma$ is the covariance, $x_r$ represents a real image, and $g$ is a generated image.

3.2 Experimental results

In the experiment, we set the parameters of the comparison experiment with the optimal parameters in order to compare the best reconstruction performance. Figures 6, 7 and 8 show the reconstructed two-dimensional super resolution MR image by using different GAN-based algorithms. Due to the small difference in visual observation, we choose a two-dimensional MRI with more prominent features and zoom in on a specific area. From Figures 6, 7, 8, and 9, when compared with other three GAN-based methods, it can be found that the FA-GAN based reconstruction algorithm has a clearer texture structure in detail than the other methods. Due to the combination of the self-attention module and channel attention module, with a richer high-frequency texture detail under a large-scale factor, the proposed FA-GAN method preserves more detailed structural information and the fine outline of the MR image. The reconstructed image has clear texture details and most of the aliasing artifacts are effectively suppressed even with 4× super resolution.

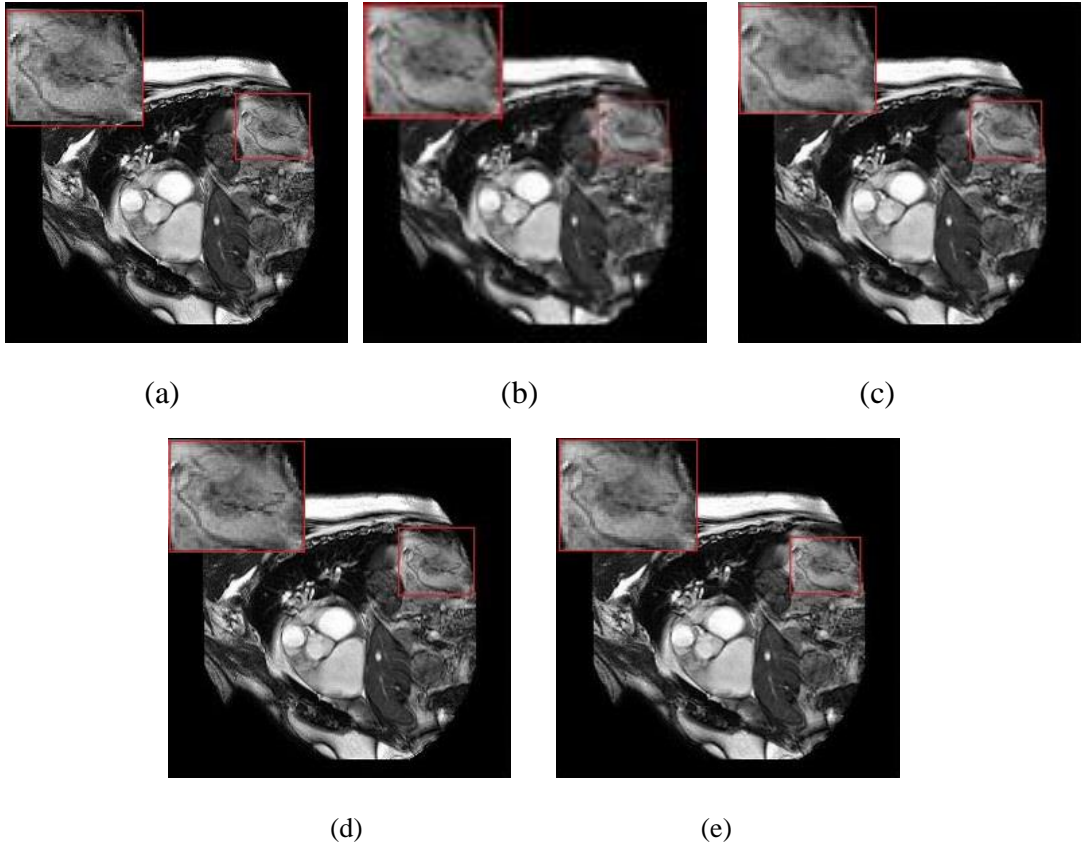

Figure 6: The reconstructed super-resolution cardiac MR images by using different GAN based methods(4×). (a) the real high resolution MR image, and the reconstructed super-resolution MRI using (b)SR-GAN(PSNR 32.37,SSIM 0.9948), (c)SA-SR-GAN(PSNR 34.07,SSIM 0.9959) ,(d)CA-SR-GAN(PSNR 34.12,SSIM 0.9963),(e)FA-GAN(PSNR 34.28,SSIM 0.9966).

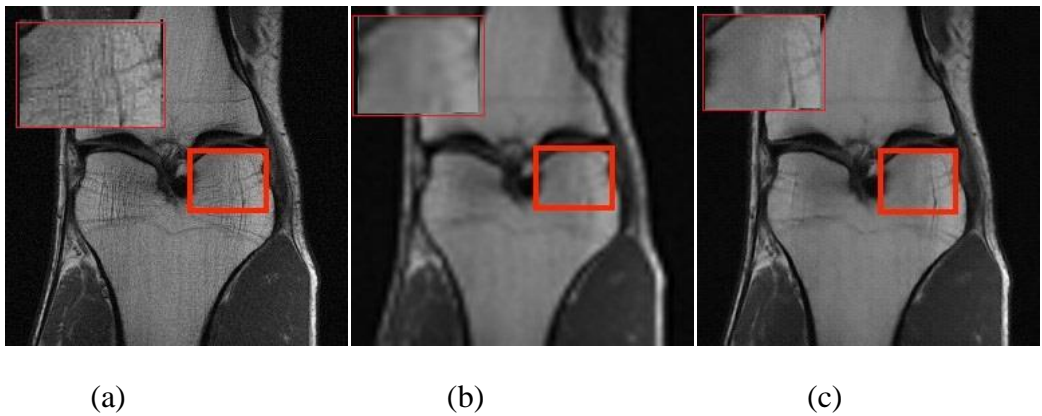

(a)          (b)          (c)

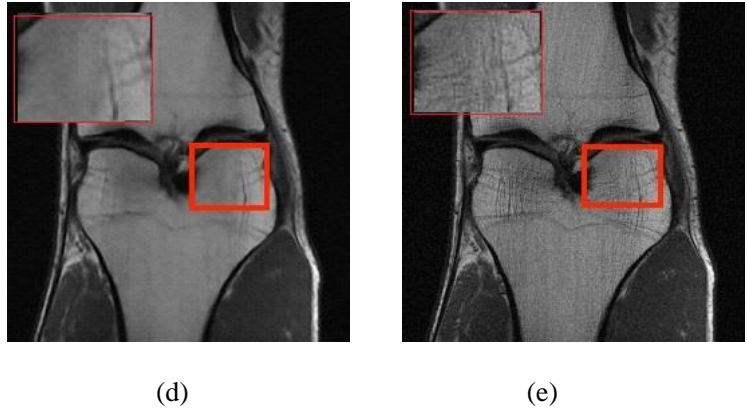

(d)          (e)

Figure 7: The reconstructed super-resolution knee MR images by using different GAN-based methods(4×). (a) the real high-resolution MR image, and the reconstructed super-resolution MRI using (b)SR-GAN(PSNR 28.90 SSIM 0.9905), (c)SA-SR-GAN(PSNR 30.10 SSIM 0.9912), (d)CA-SR-GAN(PSNR 30.05 SSIM 0.9910). (e)FA-GAN(PSNR 30.28 SSIM 0.9926).

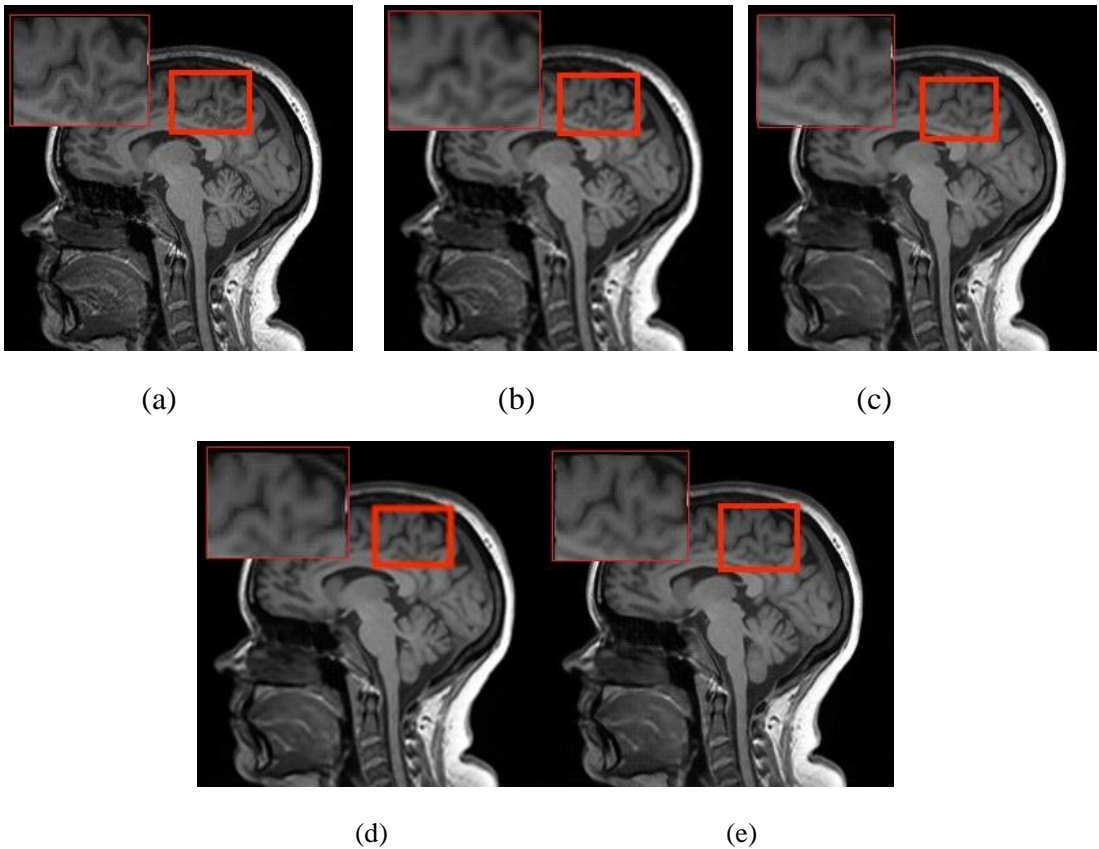

Figure 8: The reconstructed super-resolution brain MR images by using different GAN-based methods(4×). (a) the real high resolution MR image, and the reconstructed super-resolution MRI using (b)SR-GAN(PSNR 39.90 SSIM 0.9957), (c)SA-SR-GAN(PSNR 41.78 SSIM 0.9969), (d)CA-SR-GAN(PSNR 40.68 SSIM 0.9960) ,(e)FA-GAN(PSNR 42.07 SSIM 0.9974).

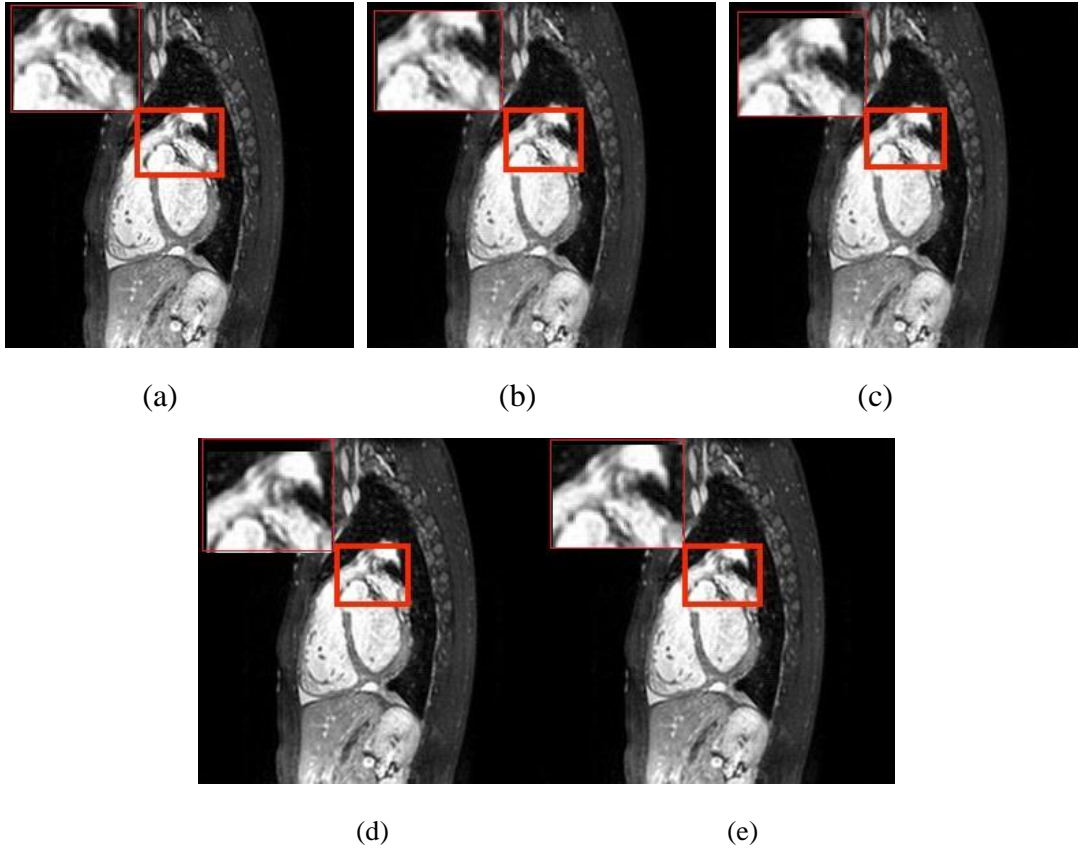

Figure 9: The reconstructed super-resolution MR images by using different GAN based methods on MM-WHS(4×). (a) the real high resolution MR image, and the reconstructed super-resolution MRI using (b)SR-GAN(PSNR 34.42 SSIM 0.9937), (c)SA-SR-GAN(PSNR 35.93 SSIM 0.9966), (d)CA-SR-GAN(PSNR 36.59 SSIM 0.9974) ,(e)FA-GAN(PSNR 36.77 SSIM 0.9980).

Table 1: Average PSNR, SSIM by using different GAN-based methods (cardiac）

| method | SRGAN | | CA-SR-GAN | | SA-SR-GAN | | FA-GAN | |
|---|---|---|---|---|---|---|---|---|
| | PSNR | SSIM | PSNR | SSIM | PSNR | SSIM | PSNR | SSIM |
| X2 | 33.73 ±3.75 | 0.9957± 0.0032 | 34.22 ±3.48 | 0.9959± 0.0028 | 35.66 ±3.23 | 0.9962± 0.0026 | **38.58 ±3.15** | **0.9979± 0.0023** |
| X4 | 32.30 ±4.26 | 0.9946± 0.0067 | 33.75 ±4.79 | 0.9953± 0.0073 | 34.09 ±4.42 | 0.9959± 0.0052 | **34.26 ±4.35** | **0.9965± 0.0058** |
| X8 | 31.35 ±4.31 | 0.9951± 0.0058 | 32.89 ±4.73 | 0.9952± 0.0078 | 33.45 ±4.13 | 0.9956± 0.0031 | **33.70 ±4.20** | **0.9960± 0.0023** |

Table 1, 2, 3 and 4 show the average values of two quantified indicators of PSNR and SSIM of 3200 two-dimensional MRI reconstructed by using different algorithms. Table 1 presents cardiac super-resolution MRI reconstruction performances by using the differences methods, Table 2 shows the brain super-resolution MRI reconstruction performances of different methods, Table 3 provides the knee super-resolution MRI reconstruction results of different methods, and Table 4 shows the MM-WHS by using different GAN-based methods. The reconstruction performances of these GAN-based methods are listed in terms of the PSNR and SSIM values at three magnifications of 2, 4, and 8 times. As shown in Tables 1-4, it can be found that the proposed FA-GAN method achieve the highest PSNR and SSIM among these four GAN-based reconstruction methods, and the following are SA-SR-GAN, CA-SR-GAN and SRGAN. From the tables, the FA-GAN method can improve the average PSNR of the reconstructed images by about 0.44-4.85 and SSIM by about 0.0003-0.0044, especially in the cardiac MR image with 2× super-resolution reconstruction. The proposed FA-GAN can improve the reconstruction performances obviously in terms of the PSNR and SSIM values.

Table 2: Average PSNR, SSIM by using different GAN-based methods（brain）

| method | SRGAN | | CA-SR-GAN | | SA-SR-GAN | | FA-GAN | |
|---|---|---|---|---|---|---|---|---|
| | PSNR | SSIM | PSNR | SSIM | PSNR | SSIM | PSNR | SSIM |
| ×2 | 43.35 ±5.46 | 0.9989± 0.0046 | 43.56 ±5.28 | 0.9991± 0.0042 | 43.97 ±5.79 | 0.9991± 0.0037 | **44.11 ±5.11** | **0.9992± 0.0039** |
| ×4 | 39.88 ±5.86 | 0.9955± 0.0068 | 40.56 ±5.12 | 0.9963± 0.0082 | 41.76 ±4.36 | 0.9970± 0.0055 | **42.11 ±4.12** | **0.9974± 0.0062** |
| ×8 | 30.83 ±5.41 | 0.8891± 0.0071 | 30.96 ±5.67 | 0.8902± 0.0093 | 31.02 ±5.32 | 0.8907± 0.0084 | **31.27 ±5.36** | **0.8935± 0.0059** |

Table 3: Average PSNR, SSIM by using different GAN-based methods (knee)

| method | SRGAN | | CA-SR-GAN | | SA-SR-GAN | | FA-GAN | |
|---|---|---|---|---|---|---|---|---|
| | PSNR | SSIM | PSNR | SSIM | PSNR | SSIM | PSNR | SSIM |
| ×2 | 34.21 ±6.51 | 0.9963± 0.0037 | 35.28 ±5.76 | 0.9964± 0.0031 | 36.28 ±5.69 | 0.9966± 0.0042 | **36.58** ±5.17 | **0.9969**± 0.0033 |
| ×4 | 28.89 ±5.81 | 0.9903± 0.0057 | 29.75 ±5.72 | 0.9908± 0.0066 | 30.13 ±5.42 | 0.9913± 0.0072 | **30.27** ±5.61 | **0.9926**± 0.0056 |
| ×8 | 26.31 ±5.29 | 0.9722± 0.0089 | 27.46 ±5.10 | 0.9813± 0.0073 | 28.22 ±4.52 | 0.9842± 0.0075 | **28.83** ±4.36 | **0.9877**± 0.0063 |

Table 4: Average PSNR, SSIM by using different GAN-based methods (MM-WHS).

| method | SRGAN | | CA-SR-GAN | | SA-SR-GAN | | FA-GAN | |
|---|---|---|---|---|---|---|---|---|
| | PSNR | SSIM | PSNR | SSIM | PSNR | SSIM | PSNR | SSIM |
| ×2 | 36.27 ±3.27 | 0.9967± 0.0032 | 38.93 ±4.16 | 0.9981± 0.0041 | 39.68 ±4.09 | 0.9989± 0.0037 | **39.78** **±3.89** | **0.9992**± **0.0034** |
| ×4 | 34.39 ±5.23 | 0.9933± 0.0031 | 35.94 ±5.49 | 0.9968± 0.0052 | 36.63 ±5.43 | 0.9973± 0.0047 | **36.81** **±5.06** | **0.9981**± **0.0027** |
| ×8 | 27.44 ±4.28 | 0.9722± 0.0046 | 27.89 ±4.36 | 0.9842± 0.0039 | 28.43 ±4.49 | 0.9863± 0.0053 | **28.92** **±4.21** | **0.9879**± **0.0042** |

Table 5 illustrates the super-resolution MRI reconstruction performances by using different GAN-based methods in terms of FID. As shown in Table 5, it can be found that the FA-GAN can effectively reduce the FID. A lower FID means that the reconstructed SR MR images are closer to the real high resolution MR images, which means that the quality of the reconstructed SR MR images is higher.

Table 5: Average FID under different methods (4×).

| method FID | SRGAN | CA-SR-GAN | SA-SR-GAN | FA-GAN |
|---|---|---|---|---|
| cardiac | 35.78±2.58 | 26.65±2.49 | 24.23±2.32 | **18.97±2.13** |
| brain | 20.38±3.06 | 17.22±3.19 | 16.57±2.86 | **12.43±2.36** |
| knee | 43.87±4.21 | 39.89±4.34 | 38.43±3.71 | **33.59±3.79** |
| MMWHS | 35.32±3.47 | 34.77±3.59 | 32.55±2.96 | **28.15±2.98** |

## 4. Discussion

To demonstrate the effect of each component, we carry out seven ablation experiments of local feature fusion block (LFFB), channel attention (CA), and self-attention (SA). By removing the local features fusion block, our model falls back to a network similar to SRGAN but with the attention block. The results confirm that making full use of local features fusion block will significantly improve performance. One possible reason is that fusing hierarchical features improves the information flow and eases the difficulty of training. We can conclude from Table 6 that the proposed FA-GAN model with all components achieves the best performance. The integration of local feature fusion block and global feature fusion block not only improves 1-2dB on PSNR, but also gets much better visual effects in image details than the other methods with part components, as shown in Figures 6, 7, 8, and 9.

According to the results of the ablation experiment, as shown in Table 6, it can be seen that the CA and LFFB modules together plays the most important role in the super resolution MR image reconstruction, which affect the reconstruction performances obviously. However, the affection of the SA module is relative small, and the reconstruction quality drops slightly. Table 7 illustrates the reconstruction effect under different connection modes. It can be clearly seen that the weighted connection has achieved better results.Thus, we use a weighted connection in our method.

Table 6: Ablation studies on MICCAI 2013 grand challenge public data set.

| Model | | ×2 | ×4 | ×8 |
|---|---|---|---|---|
| FA-GAN | PSNR | **38.58** | 34.26 | **33.70** |
| | SSIM | **0.9979** | 0.9965 | **0.9960** |
| -SA | PSNR | 37.99 | **34.29** | 33.59 |
| | SSIM | 0.9972 | **0.9966** | 0.9954 |
| -CA | PSNR | 37.84 | 34.02 | 33.42 |
| | SSIM | 0.9970 | 0.9962 | 0.9951 |
| -LFFB | PSNR | 38.04 | 34.13 | 33.65 |
| | SSIM | 0.9973 | 0.9959 | 0.9957 |
| -SA-CA | PSNR | 37.45 | 33.70 | 33.47 |
| | SSIM | 0.9968 | 0.9957 | 0.9952 |
| -SA-LFFB | PSNR | 37.27 | 33.65 | 32.89 |
| | SSIM | 0.9963 | 0.9956 | 0.9943 |
| -CA-LFFB | PSNR | 37.03 | 33.44 | 32.12 |
| | SSIM | 0.9960 | 0.9954 | 0.9938 |

Table 7: Average PSNR,SSIM under different connections(4×).

| method | Direct Connection | | Weighted Connection | |
|---|---|---|---|---|
| | PSNR | SSIM | PSNR | SSIM |
| cardiac | 33.98±4.54 | 0.9962±0.0074 | **34.26±4.35** | **0.9965±0.0058** |
| brain | 41.70±4.89 | 0.9969±0.0059 | **42.11±4.12** | **0.9974±0.0062** |
| knee | 29.89±5.45 | 0.9924±0.0063 | **30.27±5.61** | **0.9926±0.0056** |
| MMWHS | 35.67±5.99 | 0.9975±0.0032 | **36.81±5.06** | **0.9981±0.0027** |

For the selection of parameters α and β, we have done the following three sets of comparative experiments. As shown in Table 8, the experimental results show that the parameters are the optimize values when α=0.5 and β=0.5.

Table 8: Average PSNR,SSIM under different values of the parameters(4×).

| | α=0.4,β=0.6 | | α=0.5,β=0.5 | | α=0.6,β=0.4 | |
| --- | --- | --- | --- | --- | --- | --- |
| | PSNR | SSIM | PSNR | SSIM | PSNR | SSIM |
| cardiac | 33.92±4.37 | 0.9961±0.0088 | **34.26±4.35** | **0.9965±0.0058** | 33.10±4.91 | 0.9961±0.0079 |
| brain | 41.88±4.91 | 0.9966±0.0067 | **42.11±4.12** | **0.9974±0.0062** | 42.02±4.96 | 0.9970±0.0082 |
| knee | 29.73±6.25 | 0.9922±0.0089 | **30.27±5.61** | **0.9926±0.0056** | 30.13±5.84 | 0.9920±0.0077 |

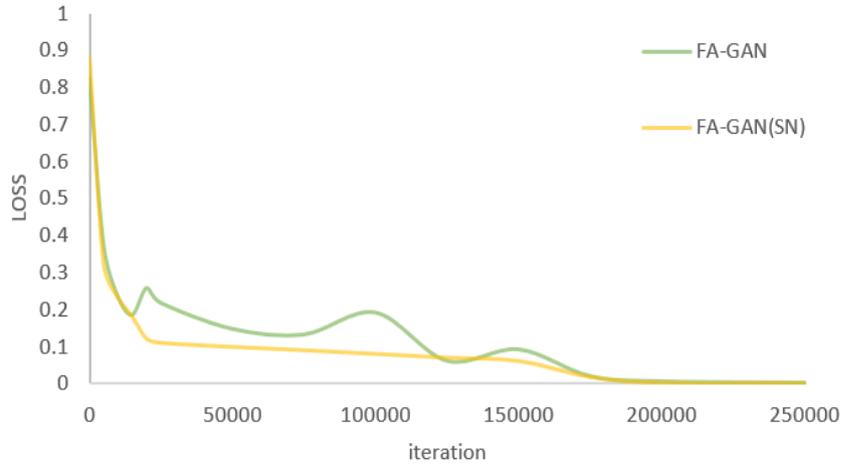

Figure 11: Comparison of loss value of FA-GAN with or without SN.

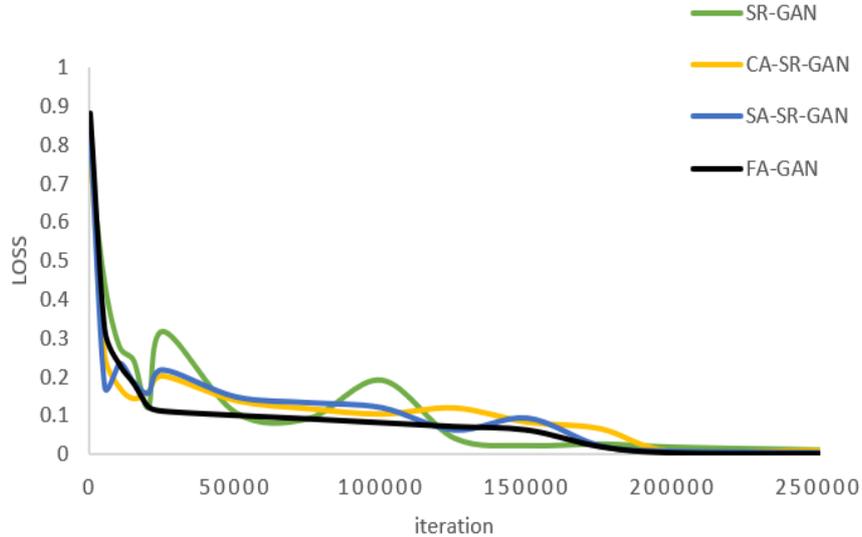

Figure 12: Comparison of loss under factor 4 and different methods.

In this paper, the spectral normalization (SN) is introduced to the discriminator network, so as to stabilize the training of GAN and limit the Lipschitz constant of the discriminator. Compared with other normalization techniques, spectral normalization does not require additional hyper parameter adjustments (set the spectral norm of all weight layers to 1). Figure 11 shows the effect of SN on FA-GAN, which makes the loss value steadily drop and makes the whole training process more stable.

Figure 12 illustrates the loss value of the training process of SR image reconstruction by using four different GAN-based methods with ×4 times. It can be found that the loss value by using FA-GAN method decreases monotonously with iteration increasing, while the other methods decrease in waves, which indicates that the proposed FA-GAN method combined with spectral normalization makes the training more stable.

## 5. Conclusion

This paper proposed a new method for super-resolution magnetic resonance images reconstruction by using fusion attention based generative adversarial networks (FA-GAN). Two different attention mechanisms are integrated into the SRGAN framework to obtain important features. Compared with the SRGAN framework, the proposed FA-GAN method can reconstruct super-resolution images with higher PSNR, SSIM and lower FID, and the reconstructed SR images preserve much closer image details to the real high-resolution image. In the future work, the proposed FA-GAN method can be used to reconstruct the super resolution MR images like 7T resolution from 3T MR equipment, which can improve the resolution of the MR image without change the hardware.

**Acknowledgement**

This work is supported in part by the National Natural Science Foundation of China (61672466, 62011530130 and 61771080), in part by Joint Fund of Zhejiang Provincial Natural Science Foundation(LSZ19F010001), in part by the Key Research and Development Program of Zhejiang Province (2020C03060) , in part by the 521 Talents project of Zhejiang Sci-Tech University, in part by the British Heart Foundation (Project Number: TG/18/5/34111, PG/16/78/32402), in part by the Hangzhou Economic and Technological Development Area Strategical Grant